\documentclass[a4paper,11pt]{article}
\usepackage[margin=1in]{geometry}
\usepackage{array}
\usepackage{caption}
\usepackage{subcaption}

\usepackage[english]{babel}
\usepackage[utf8]{inputenc}
\usepackage{asymptote}
\usepackage{amssymb,amsmath,amsthm,empheq}
\usepackage{mathtools}
\usepackage{esdiff}
\usepackage{physics}

\usepackage{graphicx}
\usepackage{subcaption}
\usepackage{tikz}%
\usepackage{tikz-3dplot}
\usepackage{xintexpr}
\usepackage{pgfplots}
\pgfplotsset{compat=1.16}
\usepgfplotslibrary{fillbetween} 
\usepgfplotslibrary{groupplots}
\usepackage{algorithm}

\usepackage{lipsum}

\newtheorem{theorem}{Theorem}[section]

\makeatletter
        \pgfdeclareplotmark{dot}
        {%
            \fill circle [x radius=0.08, y radius=0.32];
        }%
\makeatother

\DeclarePairedDelimiterX\PH[1](){
   
   #1
}

\newcommand{\E}[1]{ \mathbb{E}[#1]}

\NewDocumentCommand{\set}{o m}{%
  \IfNoValueTF{#1}
    {\{#2\}}
    {\{#1 \mid #2\}}%
}

\newcommand{\CreateRandomWalkTable}[4]{%
\xdef#4{(0,0,0)}
\xdef\oldy{2.25}
\foreach \X in {1,...,#1}
{\pgfmathsetmacro{\myx}{#2*\X}
\pgfmathsetmacro{\myy}{\oldy+1.5*#2+rand*#3}
\xdef#4{#4 (\myx,\myy,0)}
\xdef\oldy{\myy}
}}

\newcommand{\CreateRandomWalkTableB}[4]{%
\xdef#4{(0,0,0)}
\xdef\oldy{2.25}
\foreach \X in {1,...,#1}
{\pgfmathsetmacro{\myx}{#2*\X}
\pgfmathsetmacro{\myy}{\oldy+3*#2+rand*#3}
\xdef#4{#4 (\myx,\myy,0)}
\xdef\oldy{\myy}
}}

\pgfmathsetseed{1}
\CreateRandomWalkTable{140}{0.02}{0.2}{\mytabone}
\pgfmathsetseed{11}
\CreateRandomWalkTable{140}{0.02}{0.2}{\mytabtwo}
\pgfmathsetseed{17}
\CreateRandomWalkTable{140}{0.02}{0.2}{\mytabthree}
\pgfmathsetseed{32}
\CreateRandomWalkTable{140}{0.02}{0.2}{\mytabfour}

\pgfmathsetseed{1}
\CreateRandomWalkTableB{140}{0.02}{0.2}{\mytaboneB}
\pgfmathsetseed{11}
\CreateRandomWalkTableB{140}{0.02}{0.2}{\mytabtwoB}
\pgfmathsetseed{17}
\CreateRandomWalkTableB{140}{0.02}{0.2}{\mytabthreeB}
\pgfmathsetseed{32}
\CreateRandomWalkTableB{140}{0.02}{0.2}{\mytabfourB}

\makeatletter
\newcommand{\thickhline}{%
    \noalign {\ifnum 0=`}\fi \hrule height 1pt
    \futurelet \reserved@a \@xhline
}
\newcolumntype{"}{@{\hskip\tabcolsep\vrule width 1pt\hskip\tabcolsep}}
\makeatother

\title{Gaussian Recombining Split Tree}

\author{Yury Lebedev, Arunava Banerjee
\thanks{Yury Lebedev is Doctoral Candidate, Department of Electrical and Computer Engineering, University of Florida, Gainesville, FL 32611}
\thanks{Arunava Banerjee is Associate Professor, Department of Computer and Information Science, University of Florida, Gainesville, FL 32611}
}

\date{May 1, 2024}

\begin{document}
\maketitle
\thispagestyle{empty}
\begin{abstract}
Binomial trees are widely used in the financial sector for valuing securities with early exercise characteristics, such as American stock options.  However, while effective in many scenarios, pricing options with CRR binomial trees are limited. Major limitations are volatility estimation, constant volatility assumption, subjectivity in parameter choices, and impracticality of instantaneous delta hedging. This paper presents a novel tree: Gaussian Recombining Split Tree (GRST), which is recombining and does not need log-normality or normality market assumption. GRST generates a discrete probability mass function of market data distribution, which approximates a Gaussian distribution with known parameters at any chosen time interval. GRST Mixture builds upon the GRST concept while being flexible to fit a large class of market distributions and when given a 1-D time series data and moments of distributions at each time interval, fits a Gaussian mixture with the same mixture component probabilities applied at each time interval. Gaussian Recombining Split Tree Mixture comprises several GRST tied using Gaussian mixture component probabilities at the first node. Our extensive empirical analysis shows that the option prices from the GRST align closely with the market.
\end{abstract}

\section{Introduction}

Key assumption of the Black-Scholes model is that the underlying asset's price follows a geometric Brownian motion \cite{hull93}. 
  At fixed time \( t \), a Geometric Brownian Motion \( z_0 \exp(\mu t +
    \sigma W(t) ) \) has a lognormal distribution with parameters \( (\ln
    (z_0) + \mu t) \) and \( \sigma \sqrt{t} \).
What if we have a more complicated distribution at that time, for example a mixture of some number of Gaussians? We also are interested in working with recombining binomial trees since then we can follow the usual hedging practices when pricing options. How does one come up with a tree that matches any given sampled distribution at fixed time? The Gaussian Recombining Split Tree (GRST) and GRST Mixture are our answers to the posted questions.  GRST Mixture is constructed from several GRST tied at the initial node using Gaussian mixture component probabilities, where each split tree is a recombining binomial tree that can generate a discrete probability mass function of market data distribution, which approximates a Gaussian distribution with known parameters at any chosen time interval. We delve into the constraints of GRST, examining its scope and identifying potential areas where it might fall short.

\section{Related Work}
Hull \cite{hull93} shows that tree prices converge to the Black–Scholes price with modification to the Central Limit Theorem.  Historically, it's been important to understand how this convergence process occurs. If fast convergence is a requirement, how does one modify CRR parameters to accelerate it? Generally, all binomial trees exhibit self-similarity, meaning that each node is identical in a relative sense. However, Joshi's Split tree \cite{joshi-split}, and our proposed Gaussian Recombining Split Tree (GRST) are not. So, a sequence of self-similar trees is just a selection of $p, U, D$ as a function of number of steps. If we impose risk-neutrality we immediately have a link between $p$ and $U$ and $D$. In historical evolution of binomial trees the objective has consistently guided the choice of $U_n$ and $D_n$ to ensure that the limiting tree converges to the Black–Scholes model. Our non-self-similar proposed GRST tree removes this limitation and lets the market data guide the construction. However, since $\pi_n$ constrains the mean convergence for the Black-Scholes convergent trees, we have only one degree of freedom left: the variance must converge correctly. This still leaves a lot of flexibility since we have two sequences and only one condition. We will now provide a summary of the various binomial trees that have emerged due to different selections in parameter choices.

\subsection*{CRR} 
CRR  \cite{COX1979229}  matches the first two moments in the limit.
\begin{align}
    U_n &= e^{\sigma \sqrt{\Delta T}} \\
    D_n &= e^{-\sigma \sqrt{\Delta T}}
\end{align}

\subsection*{Tian tree  } 
Tian tree \cite{RePEc:wly:jfutmk:v:13:y:1993:i:5:p:563-577} matches the first three moments exactly for all $n$ using the extra degree of freedom contrasted to CRR which only matches the first two moments in the limit when $n \rightarrow \infty$.
\begin{equation}
    \begin{aligned}
    p U_n+(1-p) D_n & =M_1 \\
    p U_n^2+(1-p) D_n^2 & =M_2 \\
    p U_n^3+(1-p) D_n^3 & =M_3
\end{aligned}
\end{equation}
where $M_1=e^{r \Delta t}, \quad M_2=e^{\left(2 r+\sigma^2\right) \Delta t}, \quad M_3=e^{\left(3 r+3 \sigma^2\right) \Delta t}$.

After solving the system of equations, Tian obtained:
\begin{align}
    r_n &= e^{r \Delta T} \\
    v_n &= e^{\sigma^2 \Delta T} \\
    U_n &= \frac{1}{2}r_n v_n \left( v_n + 1 + (v_n^2+2v_n-3)^\frac{1}{2} \right) \\
    D_n &= \frac{1}{2}r_n v_n \left( v_n + 1 - (v_n^2+2v_n-3)^\frac{1}{2} \right)
\end{align}
 Tian tree is particularly useful in scenarios requiring more precision and where the underlying asset's price distribution deviates significantly from the log-normal distribution assumed by simpler models.

\subsection*{Jarrow-Rudd (JR) tree  }
The Jarrow-Rudd tree \cite{RePEc:eee:jbfina:v:10:y:1986:i:1:p:157-161} is not risk-neutral, with a distinctive adjustment feature for skewness in the underlying asset's returns distribution. Unlike the Black-Scholes and CRR models, which assume a log-normal distribution with no skew, the JR model adjusts the up and down movement probabilities to account for skewness. The JR tree influenced us in fixing the probability $\pi=\frac{1}{2}$ throughout the tree to construct GRST.
\begin{align}
    \pi &= \frac{1}{2} \\
    \mu &= r-\frac{1}{2}\sigma^2 \\
    U_n &= e^{\mu \Delta T +\sigma \sqrt{\Delta T}} \\
    D_n &= e^{\mu \Delta T -\sigma \sqrt{\Delta T}}
\end{align}

\subsection*{Trigeorgis tree  } 
The Trigeorgis tree \cite{Trigeorgis_1991} is created to match the first two moments of the log process. The author chooses the up and down moves to be symmetric in log space. The Trigeorgis tree is known for its ability to incorporate the impact of multiple sources of uncertainty and to adjust for more complex option-like features in capital investment projects. Thus, it manages the market condition with large movements in stock prices better and is well suited for options on highly volatile assets.
\begin{align}
    K &= \sigma^2 \Delta T \\
    m &= (r^2-\frac{1}{2}\sigma^2)\Delta T \\
    H &= \sqrt{K + (mK)^2} \\
    p &= \frac{1}{2}(1+m\frac{K}{H}) \\
    u &= e^H \\
    d &= e^{-H}
\end{align}

\subsection*{Joshi's Split tree } \label{jo-split-tree}
The Joshi split tree \cite{joshi-splitoriginal}  is developed by Mark Joshi \cite{joshi-splitoriginal}. The split tree model introduces a novel approach to handle the skew and kurtosis of the underlying asset's return distribution more effectively. This approach allows the model to better capture the dynamics of financial markets, particularly the leptokurtic (fat-tailed) nature of asset return distributions, which is often observed in practice. Joshi's Split tree was developed to enhance pricing options' accuracy and efficiency and introduces a half-point split to follow a time-dependent real-world drift. In the second half, it is zero; in the first half, it is a constant chosen to make the center of the tree in log space at first maturity the strike price. For binomial models, the price $C^n$ of a call and $C^{B S}$ is the price in Black-Scholes are related via 
\begin{equation}
    C^n=C^{B S}+\frac{c_n}{n}+O\left(n^{-3 / 2}\right)
\end{equation}
as summarized by Leduc \cite{leduc-split}, where $c_n$ is bounded but not constant. When $c_n$ is constant, Richardson extrapolation is used to achieve convergence at speed of order $n^{-3 / 2}$, as the author points out. For American options, no closed-form formula exists, and thus, for error expansion below, we do not know the coefficients $c_i(n)$, where $i_0$ is an arbitrary value.
\begin{equation}
    C^n=C^{B S}+\frac{c_1(n)}{n}+\frac{c_2(n)}{n^{3 / 2}}+\cdots+O\left(n^{-\left(i_0+1\right) / 2}\right)
\end{equation}

So Joshi's Split tree is a two-phase binomial tree designed to minimize oscillations of the error above with the Black-Scholes model where a drift parameter $\lambda$ is used in the binomial model up to a split time $\tau$, after which the tree becomes a CRR tree \cite{joshi-splitoriginal}. 
In Joshi's model, the time for the split is determined as the earliest time step, $\tau$, which is equal to or exceeds half the total maturity period and the drift parameter, $\lambda$, to ensure that following the split, an equal number of nodes in the tree fall on either side of the strike price $K$.

\subsection*{GARCH Option Pricing} \label{garch-modelling}

Constant volatility assumption is a key feature of the Black-Scholes model, i.e., the volatility of the underlying asset, which is a crucial determinant of the option's value, remains the same throughout the option's life. In practical terms, this means that the future volatility of the asset's returns is predicted to be the same as its historical volatility without any fluctuations over time. Actual market conditions exhibit varying levels of volatility over time due to various economic and financial factors. Thus, the more accurately one can forecast future market volatility, the more effectively one can price a derivative, and this chapter will attempt to do precisely that. Historical data is a valuable resource for estimating current and future levels of market volatility. One can identify patterns and anomalies that may influence future market movements by analyzing past market behaviors, trends, and fluctuations. Statistical methods like standard deviation and variance and more complex models such as the GARCH (Generalized Autoregressive Conditional Heteroskedasticity) model, which can extrapolate historical volatility to predict future trends are historically have been used. 

Partly due to the ability to express the likelihood function of asset returns in closed form using observed data, the family of GARCH volatility models has gained widespread application in empirical asset pricing. The maximum likelihood estimation (MLE) method can determine the model parameters, which tends to be quite challenging for most stochastic volatility models. Duan \cite{duan-garc}, inspired by the ability of GARCH to fit asset returns, proposed LRNVR (Locally Risk-Neutral Valuation Relationship) to price SPX options. Payoff function for an option, which is $max(S-K,0)$ for a call and $max(K-S,0)$ for a put where $S, K$ are stock price and strike prices respectively.
To determine the current value of options, one approach is to simulate every potential price of the underlying asset at time T, calculate the mean of the payoff function from these simulations, and then multiply this average by $e^{-rT}$ using the risk-neutral measure. So, following Hull \cite{hull93} under the risk-neutral pricing measure $Q$, the values of Call and Put are:
\begin{align*}
    Call &= e^{-rT}E_Q \left[ max(S - K,0) \right] \\
    Put &= e^{-rT}E_Q \left[ max(K - S,0) \right]
\end{align*}

We model the asset price $S_t$ as a discrete-time stochastic process, and Duan \cite{duan-garc} suggests that under the physical measure $P$, asset returns adhere to a conditional log-normal distribution:
\begin{equation} \label{duan-process}
    \ln \frac{S_t}{S_{t-1}}=r-\frac{1}{2} h_t+\lambda \sqrt{h_t}+\epsilon_t
\end{equation}
where $r$ is one-period risk-free interest rate, $\lambda$ is the asset risk premium, and $\epsilon_t$ follows a GARCH(p,q) process with zero mean and conditional variance $h_t$:
\begin{align}
    \epsilon_t \mid \phi_{t-1} \sim N\left(0, h_t\right) \text { under measure } P \text {, } \\
    h_t=\alpha_0+\sum_{i=1}^q \alpha_i \epsilon_{t-i}^2+\sum_{j=1}^p \beta_j h_{t-j} \label{eq:duan-garc-gen}
\end{align}
where $\phi_t$ is the information that includes time $t$. $\alpha_0 \geq 0, \alpha_i \geq 0 \text { for } i=1,2, \ldots, q$ and $\beta_j \geq 0 \text { for } j=1,2, \ldots, p$. For GARCH(1,1), Duan's \cite{duan-garc} equation \ref{eq:duan-garc-gen} simplifies to
\[
h_t=\alpha_0+\alpha_1 \epsilon_{t-1}^2+\beta_1 h_{t-1} 
\]
To address the heteroskedastic nature of the asset returns process in \ref{duan-process}, Duan introduced the LRNVR. Under the LRNVR framework, the anticipated return is expected to equal the risk-free rate, and the variance projected one step ahead should remain equal across both measures. Particularly within the LRNVR framework, the behavior of asset returns under the risk-neutral pricing measure Q takes the following form:
\begin{align} \label{eq:duan-simulate} 
\ln \frac{S_t}{S_{t-1}} &= r-\frac{1}{2} h_t+\xi_t, \quad \xi_t \mid \phi_{t-1} \sim N\left(0, h_t\right) \quad \text { under measure } Q, \\
h_t &= \alpha_0+\alpha_1\left(\xi_{t-1}-\lambda \sqrt{h_{t-1}}\right)^2+\beta_1 h_{t-1}.
\end{align}
Now, one can Monte-Carlo simulate asset price paths using Duan equation \ref{eq:duan-simulate}. First, market data must be used to fit model parameters, similar to GARCH parameters, and Duan's \cite{duan-garc} uses Maximum likelihood estimation as well with the following objective function:
\[
-\frac{T}{2} \ln (2 \pi)-\frac{1}{2} \sum_{t=1}^T\left\{\ln \left(h_t\right)+\left[\ln \left(S_t / S_{t-1}\right)-r-\lambda \sqrt{h_t}+\frac{1}{2} h_t\right]^2 / h_t\right\}
\]
Since its introduction, Duan's LRNVR has become a popular tool for option pricing within the GARCH framework but, as noted by Barone-Adesi \cite{barone}, empirical findings indicated that the constraints in Duan's LRNVR resulted in relatively weak performances in pricing and hedging, underperforming CBOE VIX by about 10\%.

\subsection*{Option Pricing with Markov Non-Recombining Tree}

We now review a Markov Tree (MT) model, created by Bhat and Kumar \cite{bhat1} where the behavior of the underlying asset is captured through a non-independent and identically distributed process. Closely following the authors, we assume that $S_n$ is the stock's price at time step $n$. Bhat and Kumar initially ($n=0$) start with one step of the standard binomial tree
\begin{equation}
    \begin{aligned}
& P\left(S_1=u S_0\right)=q \\
& P\left(S_1=d S_0\right)=1-q .
\end{aligned}
\end{equation}
When $n \geq 1$, the authors define two events:
\begin{equation}
    \begin{aligned}
& S_n^{+}=\left\{S_n \geq S_{n-1}\right\} \\
& S_n^{-}=\left\{S_n<S_{n-1}\right\}
\end{aligned}
\end{equation}
So the event $S_n^{-}$ happens when the stock price decreases from previous step to the next one. Similarly, the event $S_n^{+}$ happens when the stock price increases from previous step.
Now, the authors \cite{bhat1} introduce their model for the evolution of $S_n$:
\begin{equation} \label{mt-eq}
    \begin{aligned}
P\left(S_{n+1}=v S_n \mid S_n^{+}\right) & =q^{+} \\
P\left(S_{n+1}=w S_n \mid S_n^{+}\right) & =1-q^{+} \\
P\left(S_{n+1}=x S_n \mid S_n^{-}\right) & =q^{-} \\
P\left(S_{n+1}=y S_n \mid S_n^{-}\right) & =1-q^{-} .
\end{aligned}
\end{equation}
 In Equation \ref{mt-eq} the authors have introduced four symbols,  $v, w, x, y$, each representing distinct factors that permit changes in the stock price at every time step. Bhat and Kumar state that $q, q^{+}, q^{-}$, respectively, represent risk-neutral version of the probabilities $P(u)$, $P(u|u)$, and $P(u|d)$.
 
Let $J_n$ be the vector of possible states that the asset can be in after $n$ steps of following the Markov Tree. Bhat and Kumar \cite{bhat1} show that $J_n$ contains $n^2-n+2$ elements, which means that some of the states recombine. For the standard binomial model like CRR tree, the number of states increase linearly. The polynomial growth is crucial for the computational feasibility of the Markov tree, as it implies that the number of states does not exhibit $O(2^n)$ complexity, which is typical of a general non-recombining tree. If the asset goes up from $S_0uvw$ it will attain the same value as if it went up from $S_0uwx$.

If $r$ is the risk-free rate of interest, the authors use the first fundamental theorem of asset pricing to define risk-neutral probabilities:
\begin{equation}
    q=\frac{\exp (r \Delta t)-d}{u-d}, \quad q^{+}=\frac{\exp (r \Delta t)-w}{v-w}, \quad q^{-}=\frac{\exp (r \Delta t)-y}{x-y}
\end{equation}
Bhat and Kumar \cite{bhat1} then employ their Markov tree to price a European call option with strike $K$ and asset spot price $S_0$. For a random variable $X(\omega)$, let $X_{+}(\omega)$ be the random variable
\begin{equation}
    X_{+}(\omega)= \begin{cases}X(\omega) & X(\omega)>0 \\ 0 & X(\omega) \leq 0\end{cases}
\end{equation}
Assuming $S_N$ is the assets price at expiry, the authors employ their tree to generate a risk-neutral p.m.f. for the random variable $S_N$.  Then MT model call option price is
 \cite{bhat1}
\begin{equation}
    C=e^{-r Y} E\left[\left(S_N-K\right)_{+}\right]
\end{equation}
The authors derive the exact p.m.f generated by Markov tree, in which case the expectation can be evaluated and the option price becomes
\begin{equation}
    C=e^{-r Y} \sum_{\substack{\sigma \in J_N \\ \sigma>K}}(\sigma-K) P\left(S_N=\sigma\right)
\end{equation}
Notice, the pricing is similar as GARCH pricing where we compute the expectation of possible paths and discount back, in contrast of our proposed approach to use the tree for delta-hedging.

Bhat and Kumar \cite{BHAT2012762} follow up on their research and demonstrate that the discrete p.m.f. of log returns generated by the Markov tree is approximated by a continuous mixture of two Gaussian distributions. Finally, the authors use normal mixture distribution and risk-neutral pricing to derive a closed-form expression for European call option prices.

\section{Gaussian Recombining Split Tree Mixture}

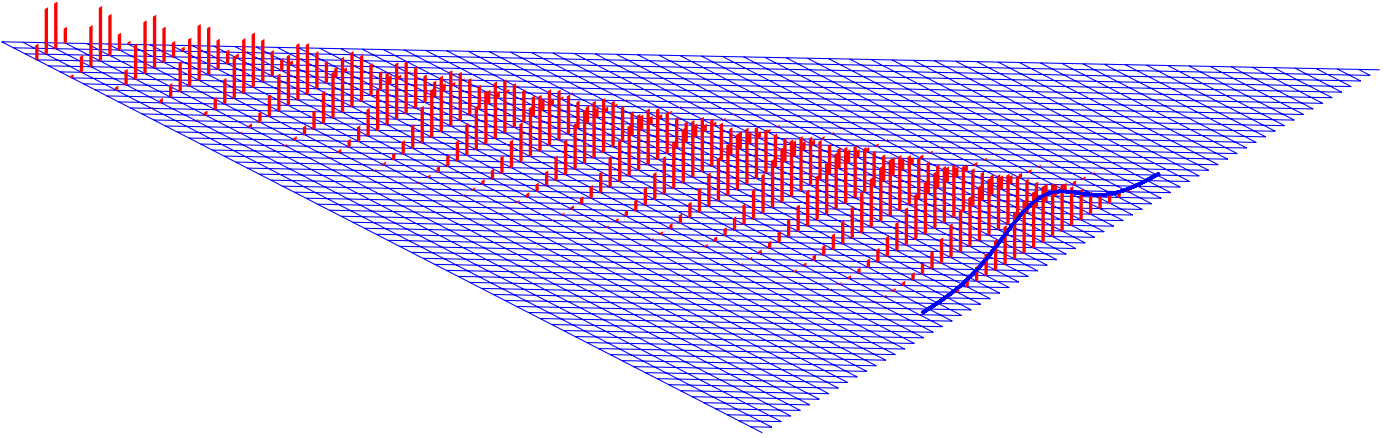
\begin{figure} [h]
    \centering

\usetikzlibrary{3d}
\usetikzlibrary{fpu}

\xintDigits := 8;
\xintdeffloatfunc binom(k, n, p) := binomial(n, k)*p^k*(1-p)^(n-k);

\tdplotsetmaincoords{110}{-30}

\begin{tikzpicture}[scale=0.25,binom tree/.style={insert path={%
  foreach \X in {1,...,#1}
 {foreach \Y in {1,...,\X} {(\X,\Y-\X/2) -- (\X+1,\Y+1/2-\X/2)
 (\X,\Y-\X/2) -- (\X+1,\Y-1/2-\X/2)} }}},tdplot_main_coords]

 \begin{scope}[canvas is xy plane at z=0]

  \draw[blue,binom tree=65];

 \end{scope}

 \tdplotsetrotatedcoords{0}{0}{0}

 \begin{scope}[tdplot_rotated_coords,declare function={gauss(\x,\y,\z)=1/(\y*sqrt(2*pi))*exp(-((\x-\z)^2)/(2*\y^2));}]

  \xintFor* #1 in {\xintSeq[3]{0}{65}}:
  {
  \begin{scope}[canvas is yz plane at x=#1+1]
   \fill[red,thick] plot[ybar] coordinates {\xintthecoords
     \xintfloatexpr subs(
           seq((y-#1/2+0.5,A*binom(y, #1, 0.5)), y=0..#1)
          ,A=4*sqrt(#1))\relax};

  \end{scope}
   \begin{scope}[canvas is yz plane at x=64,scale=0.5]
   \draw[blue,very thick] plot[variable=\x,domain=-25:25,samples=51,smooth] (\x+0.5,{105*gauss(\x,8,0)});
  \end{scope}
  }
 \end{scope}
\end{tikzpicture}
\caption{ Sampling on GRST(0) with large number of nodes approaching Normal distribution}
    \label{fig:grst-0-limit}
\end{figure}

We define a Gaussian Recombining Split Tree of order 0 or GRST(0) to be the CRR tree with the following modification. If \( S_n \) denotes an asset price at time \( t_n \) for \(
n=0,1,\ldots N \) then this price changes according to the rule
\begin{equation} \label{add-crr}
     S_{n+1} = S_n + H_{n+1}, 0 \le n \le N-1
\end{equation}
   
where \( H_{n+1} \) is a Bernoulli random variable such
that
\[
    H_{n+1} =
    \begin{cases}
        u, & \text{with probability \( p = \frac{1}{2} \)} \\
        d, & \text{with probability \( q = 1-p = \frac{1}{2} \)}.
    \end{cases}
\]
GRST fixes the probabilities: $p=q=\frac{1}{2}$. Also note that in Equation \ref{add-crr} we are intentionally adopting an additive model, leading to a Normal distribution, as opposed to a multiplicative model, which yields a Log-Normal distribution. GRST(0) employs the indexing convention where a pair of integers \( (n,j) \), with \( n=0,\ldots N \) and \( j=0,\ldots,
n \) identifies each node in the tree. The price of the underlying asset at trading time \( t_n \) is then \[ S_n = S+ju + (n-j)d. \]  The probability of
going from price \( S \) to price \( S_n \) is
\[
    p_{n,j} = \binom{n}{j} p^j (1-p)^{n-j} = \binom{n}{j} \frac{1}{2^n}.
\]
We sample GRST(0) tree through a process which starts at the root and ends at the leaf node, at each node $i$ we use  Bernoulli random variable $H_i$ to follow up node with probability $p=\frac{1}{2}$  or to follow down node with $q=1-p$ , the sampled PDF is approaching Normal distribution as displayed in Figure \ref{fig:grst-0-limit}.

\subsection*{Nearest Neighbor Lattice} 

Assume we have a process which follows a Brownian motion with starting value of $S_t$. As time progresses, the sampled Gaussian distribution becomes more dispersed so, for this process, the delta Gaussian centered at $S_t$, after step $\Delta t$ evolves to $S_{t+\Delta t}$, which is Gaussian: 
\begin{equation} 
    S_{t+\Delta t} \sim N\left( S_t+\mu\Delta t, \sigma^2 \Delta t \right)
\label{eq:GaussEvolve}    
\end{equation}

We now construct a single period GRST(0) tree. Specifically, during time $\Delta t$ from $S_t$ the asset can reach $S_t+u$ with risk-neutral probability $p$ or $S_t+d$ with risk-neutral probability $1-p$. If we embed this single period binomial tree in a lattice what is the nearest neighbor distance $l_t$ at the CRR layer, which is the last leaf layer? 

Matching mean and variance of $S_{t+\Delta t}$
\begin{empheq}[left=\empheqlbrace]{align}
p(S_t+u)+(1-p)(S_t+d) = S_t + \mu\Delta t \label{eq:one} \\
p(S_t+u)^2+(1-p)(S_t+d)^2 - (S_t + \mu \Delta t)^2 = \sigma^2\Delta t \label{eq:two}
\end{empheq}

For GRST(0), $p=0.5$, the system has the following solution

\begin{empheq}[left=\empheqlbrace]{align}
 u &=  \mu \Delta t + \sigma  \sqrt{\Delta t }\label{eq:9} \\
 d  &=  \mu \Delta t - \sigma \sqrt{\Delta t } \label{eq:8} 
\end{empheq}
GRST(0) nearest neighbor lattice distance becomes
\begin{equation} \label{nn-latice-distance}
    l_t = u-d = 2\sigma \Delta t
\end{equation}

\subsection*{GRST(1) Construction} \label{grst1_constr}

\begin{figure}[htbp]
\centering
\includegraphics[scale=0.26]{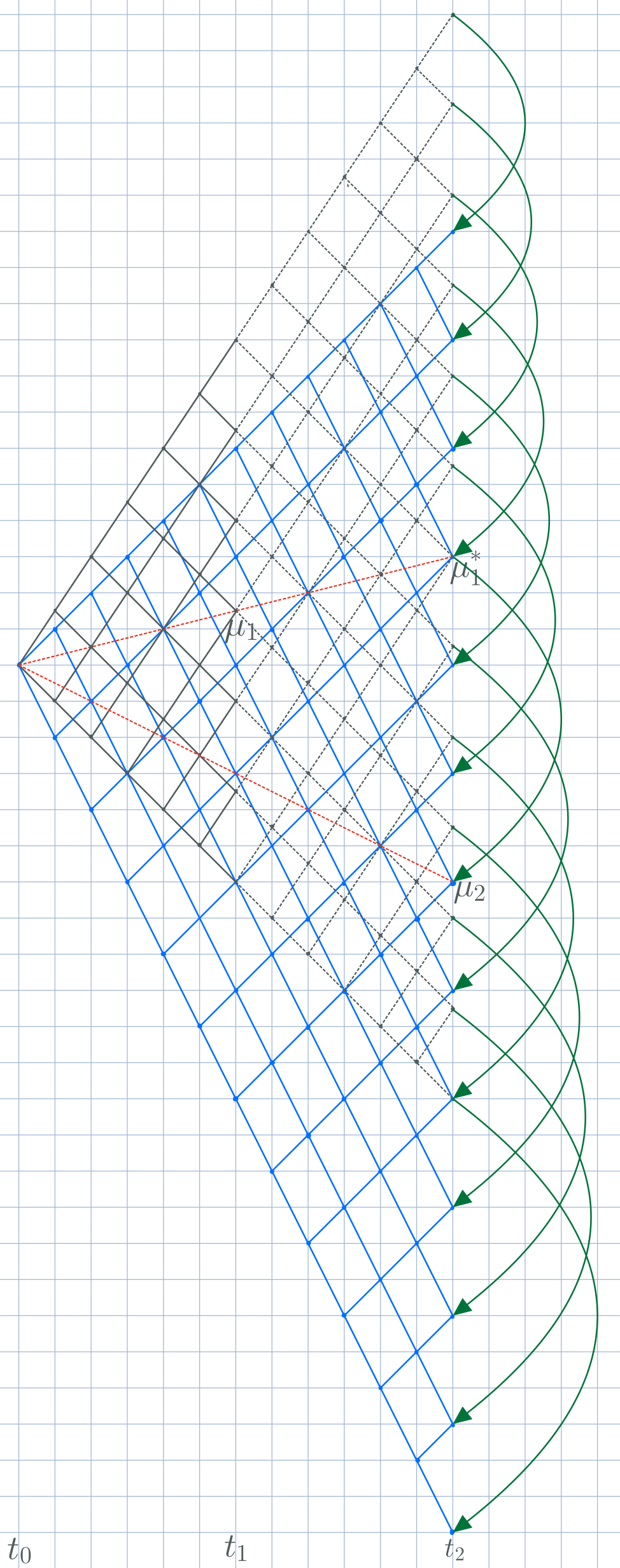}
\caption{Extension of GRST(0) $\mathcal{T}_1$ until time $t_2$ and affine transformation between nearest neighbors of $\mathcal{T}_1^*$ and $\mathcal{T}_2$ }
\label{fig:grst1-construction}
\end{figure}

We now proceed to build a recombining binomial tree which when sampled can reconstruct a Gaussian $G_1 = N(\mu_1,\sigma_1^2)$ at time $t_1$ and a Gaussian $G_2 = N(\mu_2,\sigma_2^2)$ and time $t_2$. Naively, one can attempt to use GRST(0) trees, however then two trees $\mathcal{T}_1$ and $\mathcal{T}_2$ are needed and only by chance will the nearest neighbor lattices for $\mathcal{T}_1$ and $\mathcal{T}_2$ match at time $t_1$. What if we start with GRST(0) $\mathcal{T}_1$ and then at time $t_1$ performed a split similar to a Joshi's Split Tree, such that at time $t_2$ the tree must produce the distribution $G_2$ or in other words, the nearest neighbor lattice distance at $t_2$ must be exactly same as if generated by $\mathcal{T}_2$? We quickly recognized that using constant $u$ and $d$ will not work as it was the case for GRST(0) or a Joshi's Tree and the tree parameters must be path dependent.

The lattice nearest neighbors at $t_1$ are samples from $G_1$ and similarly the lattice nearest neighbors at $t_2$ are samples from $G_2$. We extend the tree  $\mathcal{T}_1$ until time $t_2$ meaning we build this GRST(0) to have the same number of steps as $\mathcal{T}_2$ as shown in Figure \ref{fig:grst1-construction}. Extended $\mathcal{T}_1^*$ reconstructs a Gaussian $G_1^* = N(\mu_1^*,{\sigma_1^*}^2)$ at time $t_2$. However, now we have two Gaussian at time $t_2$: $G_1^*$ and $G_2$ and as a key point we can perform an affine transformation from  $G_1^*$ to $G_2$ using Theorem \ref{affine-norm-transforms} thus relating the tree parameters of $\mathcal{T}_1$ to $\mathcal{T}_2$ and effectively transforming the lattice nearest neighbors of $\mathcal{T}_1^*$ to the lattice nearest neighbors of $\mathcal{T}_2$ as displayed in Figure \ref{fig:grst1-construction} via the green links. Finally, these construction steps will allow us to compute dynamic node dependent split that we must perform at time $t_1$.

\begin{theorem}  \label{affine-norm-transforms}
Let $\vec{x} \sim N(\vec{\mu}, \Sigma)$. Then, any linear affine transformation of $\vec{x}$ is also a multivariate normally distributed:
\begin{equation}
    \vec{z}=A\vec{x}+\vec{b} \sim N\left(A \vec{\mu}+\vec{b}, A \Sigma A^{\mathrm{T}}\right)
\end{equation}
 
\end{theorem}

In a uni-variate case, affine transformation between
\begin{align}
    G_1 &= N(\mu_1,\sigma_1^2) \\
    G_2 &= N(\mu_2,\sigma_2^2)
\end{align}
is
\begin{equation} \label{affine-transform-gaussian}
    G_2 = aG_1+b
\end{equation}
where
\begin{align} \label{ab-transf}
a &= \frac{\sigma_2}{\sigma_1} \\
b &= \mu_2-\frac{\sigma_2}{\sigma_1}\mu_1
\end{align}

\begin{figure} [h]
    \centering
    \begin{subfigure}[b]{0.41\textwidth}
        \includegraphics[scale=0.29]{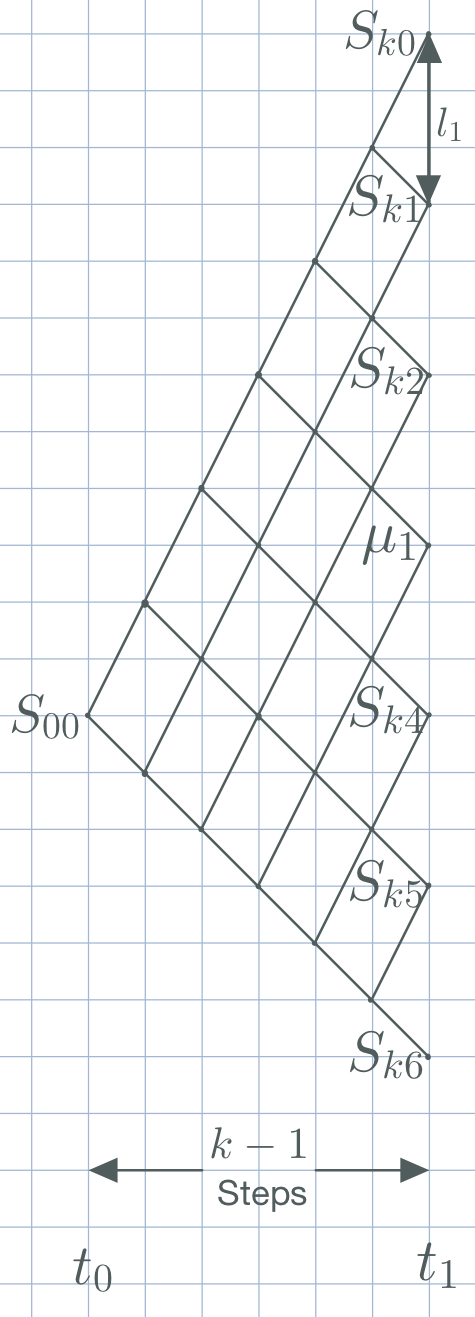}{a}
    \end{subfigure}
    \begin{subfigure}[b]{0.41\textwidth}
    	\includegraphics[scale=0.29]{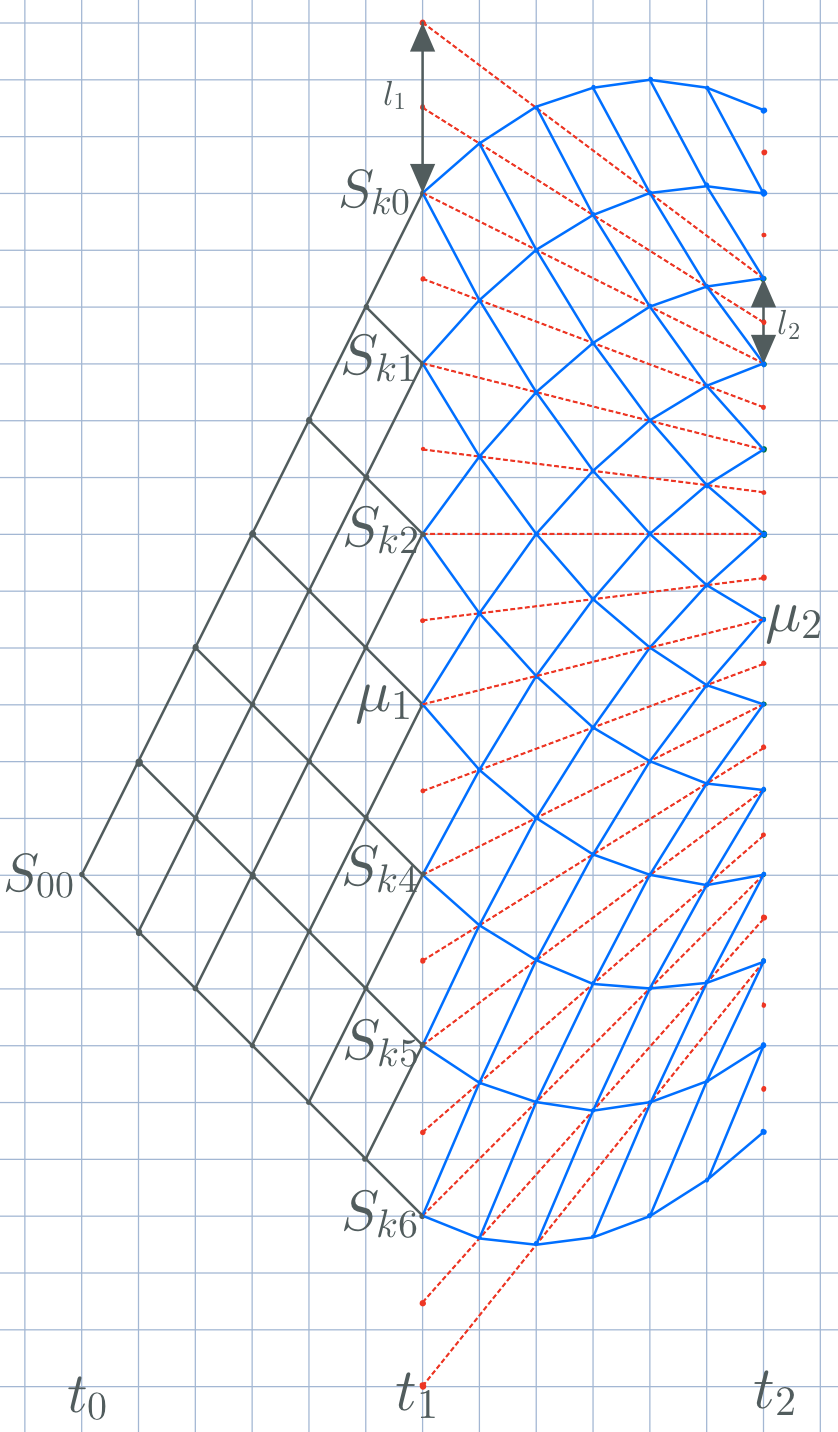}{b}
    \end{subfigure}
    \caption{Performing $k-1$ steps on GRST(0) where $k$ is odd (a) and construction of GRST(1) (b)}
    \label{fig:grst1-final}
\end{figure}

Let $S_{0,0}$ be the root of the GRST(0) $\mathcal{T}_1$ at time $t_0$ and assume, for simplicity of the implementation, we perform an even number of $k-1$ steps on it, where $k$ is odd, to reach a nearest neighbor lattice of $\mathcal{T}_1$ at time $t_1$ as displayed in Figure \ref{fig:grst1-final} on the left. Since GRST(0) is a binary recombining tree, at time $t_1$ after $k-1$ steps, the nearest neighbor lattice vector $\vec{S_1}$ will contain $k$ points: $\left[ S_{k,0}, S_{k,1}, \ldots \mu_1 \ldots S_{k,k-2} S_{k,k-1}\right]$, where $\mu_1 = S_{k, \lfloor k/2 \rfloor}$. This is consistent with the delta Gaussian centered on $S_{0,0}$ at $t_0$ drifting to a Gaussian centered at $\mu_1$ at $t_1$. We extend $\mathcal{T}_1$ to time $t_2$ over the same number of even steps of even steps $k-1$ as seen in Figure \ref{fig:grst1-construction}. Extended binary recombining GRST(0) $\mathcal{T}_1^*$ will have $2(k-1)+1 = 2k-1$ points, forming an extended nearest neighbor lattice vector $\vec{S_1}^*$ = $\left[ S_{2k-1,0}, S_{2k-1,1}, \ldots \mu_1^* \ldots S_{2k-1,2k-3} S_{2k-1,2k-2}\right]$, with $\mu_1^* = S_{2k-1, \lfloor \frac{2k-1}{2} \rfloor}$.  Then according to equation \ref{affine-transform-gaussian}, the nearest neighbor lattice vector $\vec{S_2}$ is the affine transform of $\vec{S_1}^*$:
\begin{equation} \label{affine-transform-gaussian-s1}
    \vec{S_2} = a\vec{S_1}^*+b\mathbf{1}_{2k-1}
\end{equation}
where $a$ and $b$ are given by equations \ref{ab-transf} and $\mathbf{1}_{2k-1} = $ 
$\begin{bmatrix}
1 \\
\vdots \\
1 \\
\end{bmatrix}$
with $2k-1$ elements. 

This transformation expresses the nearest neighbor lattice points for $\mathcal{T}_2$ at time $t_2$ as a function of tree parameters of $\mathcal{T}_1$ and in a sense locks in the final level of the split which must be performed at time $t_1$. To continue the construction of the split, we must establish a bijective mapping between the elements of $\vec{S_2}$ and $\vec{S_1}$, but unfortunately, the dimensions of those vectors do not match. We proceed by embedding the vectors on the infinite nearest neighbor lattice with lattice distances $l_1/2$ and $l_2/2$ correspondingly and establish the bijection between the two latices by lining up the means $S_{t_1},m$ and $S_{t_2},m$. This linear lattice correspondence is marked by dotted red lines in Figure \ref{fig:grst1-final} on the right, which also shows the final constructed split tree in blue with new dimension which is time between $t_1$ and $t_2$.

\begin{figure} [h]
\centering
\includegraphics[scale=0.26]{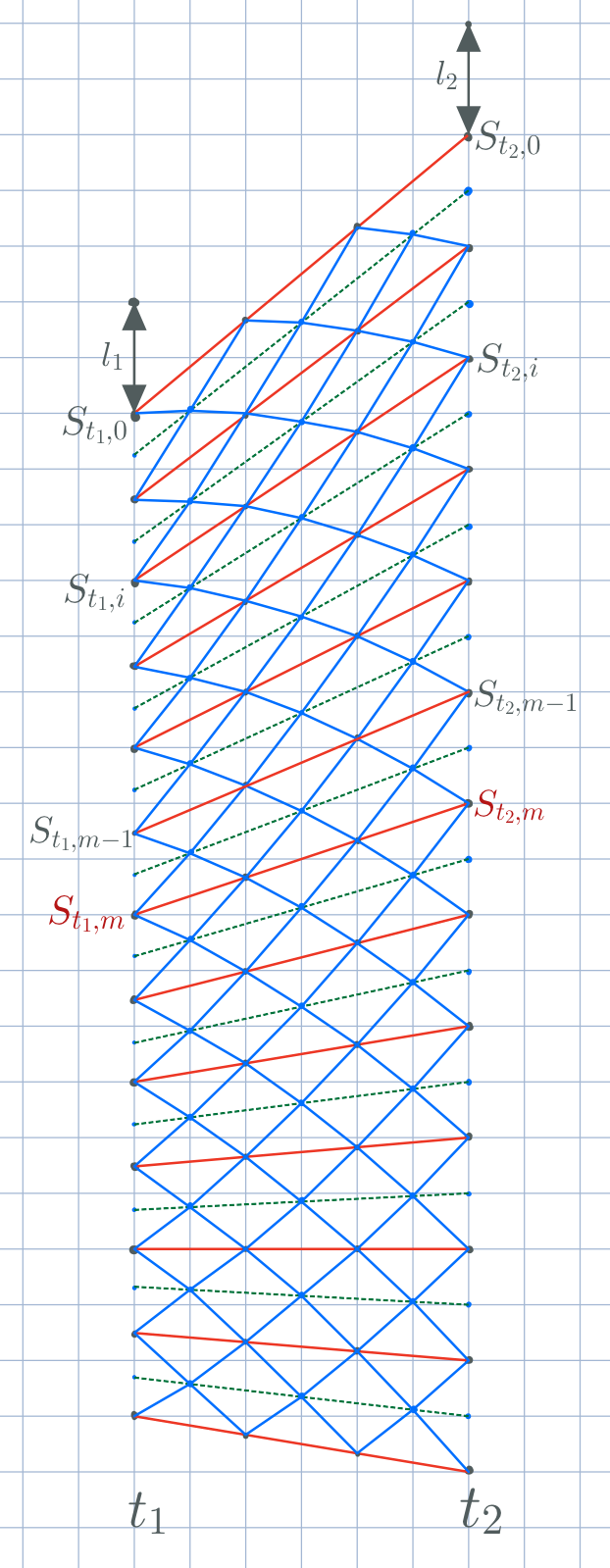}
\caption{Split tree connecting augmented GRST(0) leaf layer to augmented GRST(1) leaf layer. Red lines connect the nearest neighbor lattice points of $\mathcal{T}_1$ at $t_1$ to corresponding lattice points of $\mathcal{T}_2$ at $t_2$. Green dotted lines connect the corresponding nearest neighbor lattice midpoints between $\mathcal{T}_1$ and $\mathcal{T}_2$. }
\label{fig:grst1-lattice-flow}
\end{figure}

After the bijection is established, both augmented lattice vectors $\vec{S_1}=\vec{S_{t_1}}$ and $\vec{S_2}=\vec{S_{t_2}}$ will contain $2(k-1)+1 + 2(k-1) = 4(k-1)+1 = 4k-3$ vertices. We place these vertices on a coordinate grid with time increment step of $\Delta t = \frac{t_2-t_1}{k-1}$ as is illustrated in Figure \ref{fig:grst1-lattice-flow}, where the bijection between the augmented nearest neighbor lattice points of $\mathcal{T}_1$ at time $t_1$ and corresponding augmented lattice points of $\mathcal{T}_2$ at time $t_2$ is shown in red and the augmented midpoints are connected with the green dotted lines. The tree is now constructed by computing the vertices at each time step, which are simply the function values of red and green lines evaluated at the time steps and then connecting the edges accordingly. The recombining requirement are satisfied by construction: after each step $\Delta t$, the blue lattice points lie either on dotted green line connecting the midpoints of original nearest neighbor lattices at $t_1$ and $t_2$ or on the red lines connecting the nearest neighbor lattices at those times. For example, a red line that connects $\mu_1$ at $t_1$ to $\mu_2$ at $t_2$ will contain lattice points after each $2 \Delta t$ time steps thus guaranteeing that the tree is recombining. 

Given a vector $\vec{S_{t_1}}$ and $\vec{S_{t_2}}$ both of the same dimension $p=4k-3$ we compute the slopes of the corresponding red and dotted green lines as 
\begin{equation}
   \vec{\theta} = \frac{\vec{S_{t_2}}-\vec{S_{t_1}}}{t_2-t_1}
\end{equation}
Introduce a linear space step time vector 
\begin{equation}
    \vec{t}_{k-1}^T = [\Delta t, 2\Delta t, \ldots (k-1)\Delta t]
\end{equation}
then our nearest neighbor augmented lattice points matrix is given by
\begin{equation}
    \Lambda_{p \times k-1} = \vec{\theta} \vec{t}^T + \vec{S_{t_1}}\mathbf{1}_{k-1}^{T}
\end{equation}

\begin{figure} [htbp]
\centering
\includegraphics[scale=0.24]{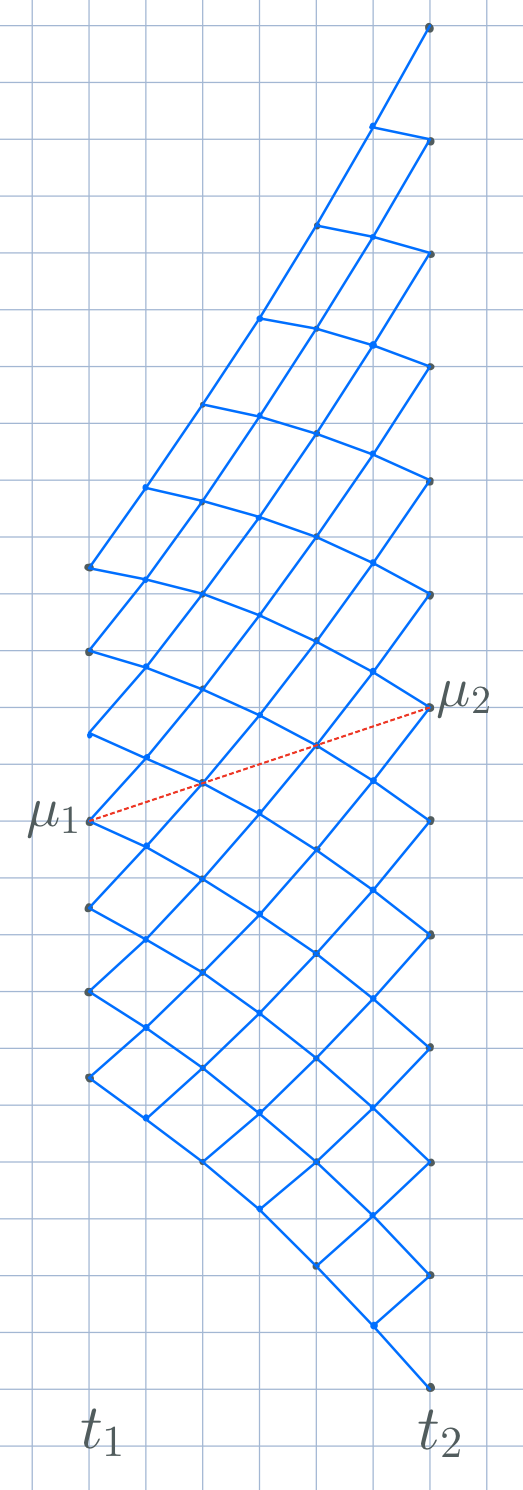}
\caption{$\Lambda$ augmented nearest neighbor matrix reduced to construct GRST(1)}
\label{fig:grst-reduced}
\end{figure}

Before the augmentation, $\vec{S_1}$, which was the last layer of initial GRST(0) tree, had $k$ elements and $\vec{S_2}$ had $2(k-1)+1=2k-1$ elements, assuming again that $k-1$ is the even number of steps taken before the split. This means the nearest neighbor augmented matrix $\Lambda_{p \times k-1}$ contains top and bottom left triangular matrices of size $\frac{4(k-1)+1-(2k-1)}{2}=k-1.$ We can now link the split layer into GRST(0) by interweaving columns from $\Lambda$ at each time step. 

We can see then $\Lambda$ augmented nearest neighbor matrix in Figure \ref{fig:grst1-lattice-flow} reduces to the data structure illustrated in Figure \ref{fig:grst-reduced}, which results in dynamic split where each blue edge is piece-wise linear but all having different slopes. 

\subsection*{GRST(N) Construction}

\begin{figure} [h]
\centering
\includegraphics[scale=0.21]{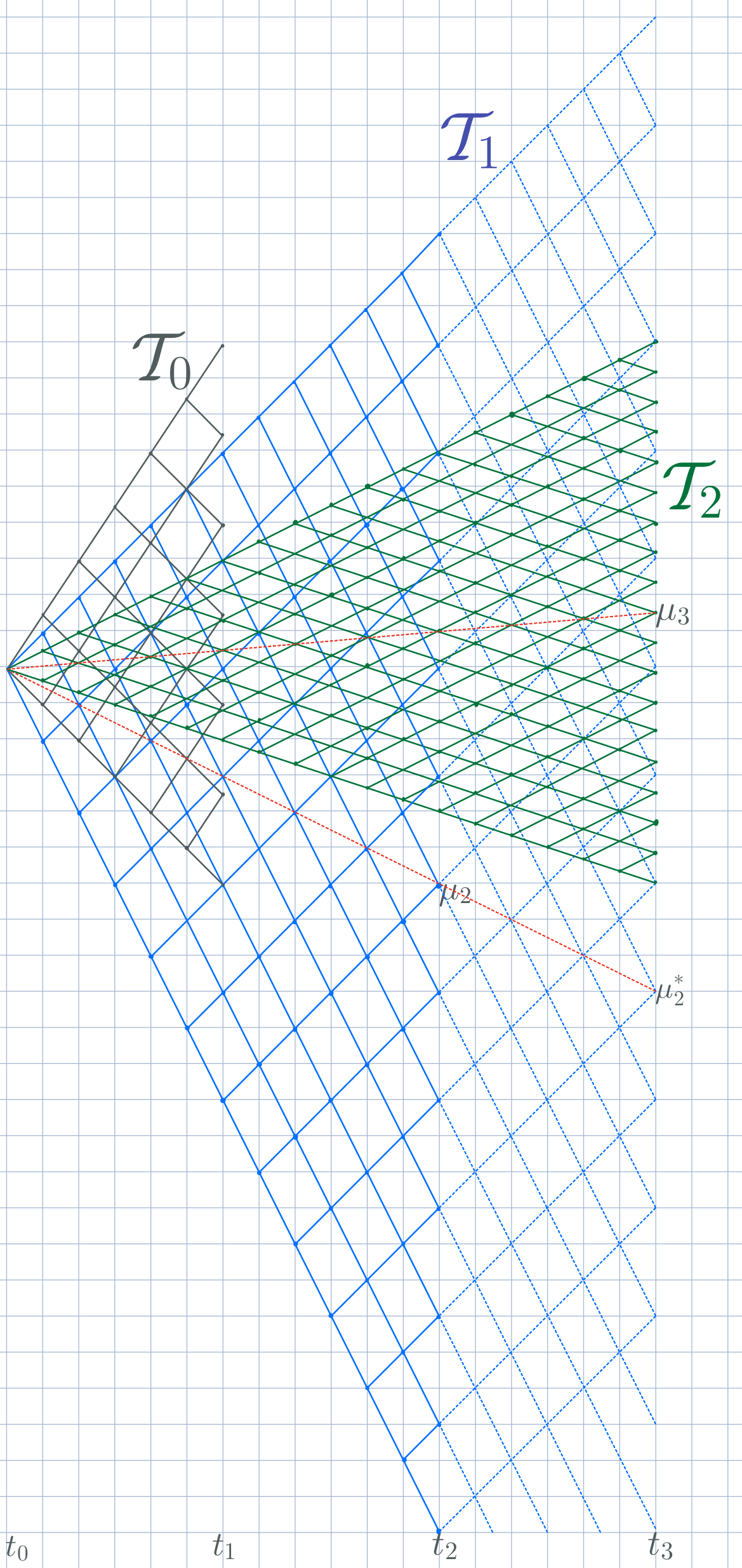}
\caption{Construction of GRST(3)}
\label{fig:grst3}
\end{figure}

We extend our construction of GRST(1) to a Gaussian Recombining Split Tree of order N. Assume we are given $N$ Gaussian: $G_1 = N(\mu_1,\sigma_1^2)$ at time $t_1$, $G_2 = N(\mu_2,\sigma_2^2)$ at time $t_2$, $\ldots$, $G_N = N(\mu_N,\sigma_N^2)$ at time $t_N$. We construct GRST(1) data structure  $\mathcal{T}_2$ to match the first two distributions as described in Section \ref{grst1_constr}, where the first split is performed at time $t_1$. We can link in the third tree by building a GRST(0) to $\mathcal{T}_2$ final layer at time $t_2$ and now perform a GRST(1) construction to time $t_3$ as shown in Figure \ref{fig:grst3} to build $\mathcal{T}_3$.

\begin{figure} [htbp]
\centering
\includegraphics[scale=0.19]{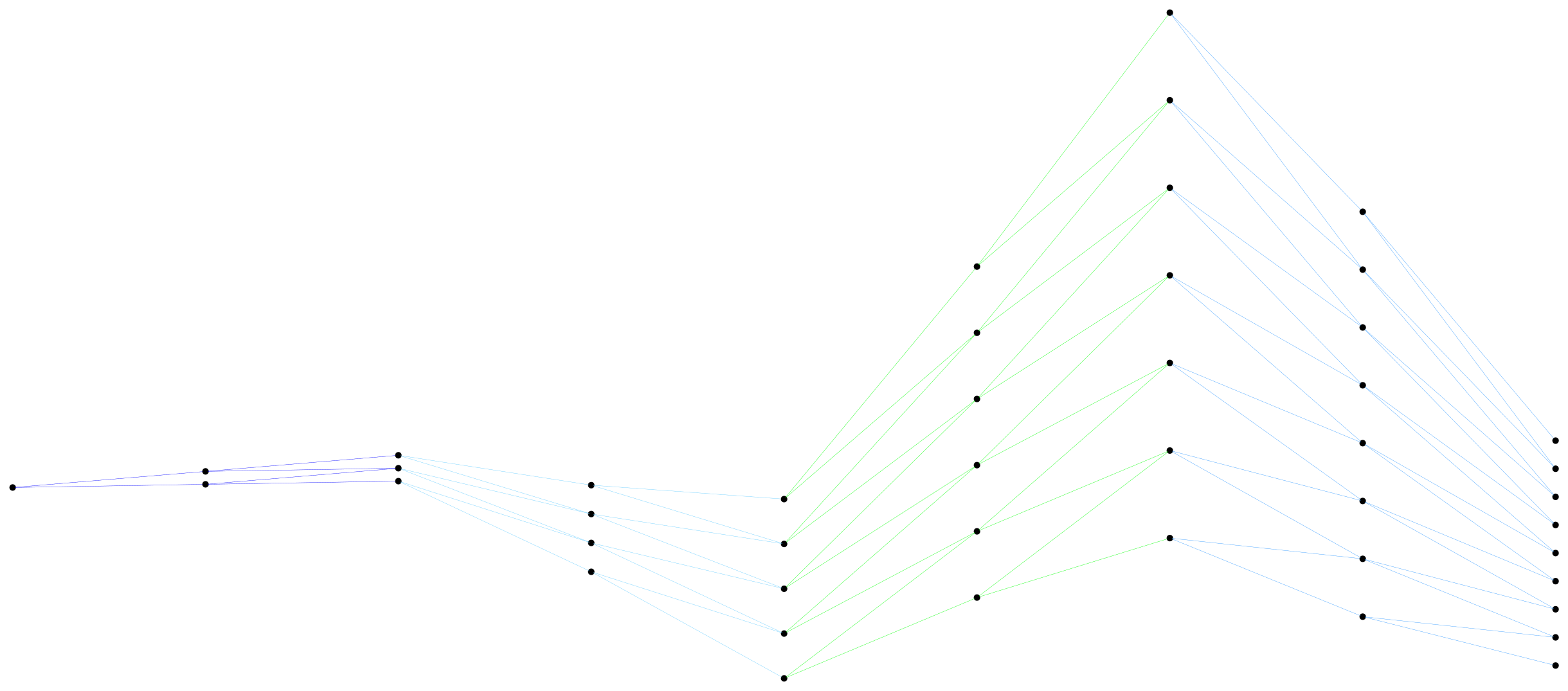}
\caption{GRST(3) with $k=3,t_1=0.1, \mu_1=1, \sigma_1=0.15; t_2=0.15, \mu_2=-3.5, \sigma_2=0.6; t_3=0.55, \mu_2=2, \sigma_2=0.75; t_4=0.68, \mu_2=-0.5, \sigma_2=0.25 $ }
\label{fig:grst3_2}
\end{figure}

We can similarly proceed recursively to construct GRST(N). Given a GRST($N-1$) at time $t_{N-1}$ represented by $\mathcal{T}_{N-1}$ data structure we construct GRST($N$) by building a GRST(0) to $\mathcal{T}_{N-1}$ final layer at time $t_{N-1}$ followed by a GRST(1) construction from  time $t_{N-1}$ to time $t_N$. Figure \ref{fig:grst3_2} displays constructed GRST(3) with matched negative means at time $t_2$ and $t_4$ demonstrating the flexibility of GRST(N) trees.  

\subsection*{Extracting Time-Series Data} \label{extract-time-series}

\begin{figure} [h]
\centering
\includegraphics[scale=0.21]{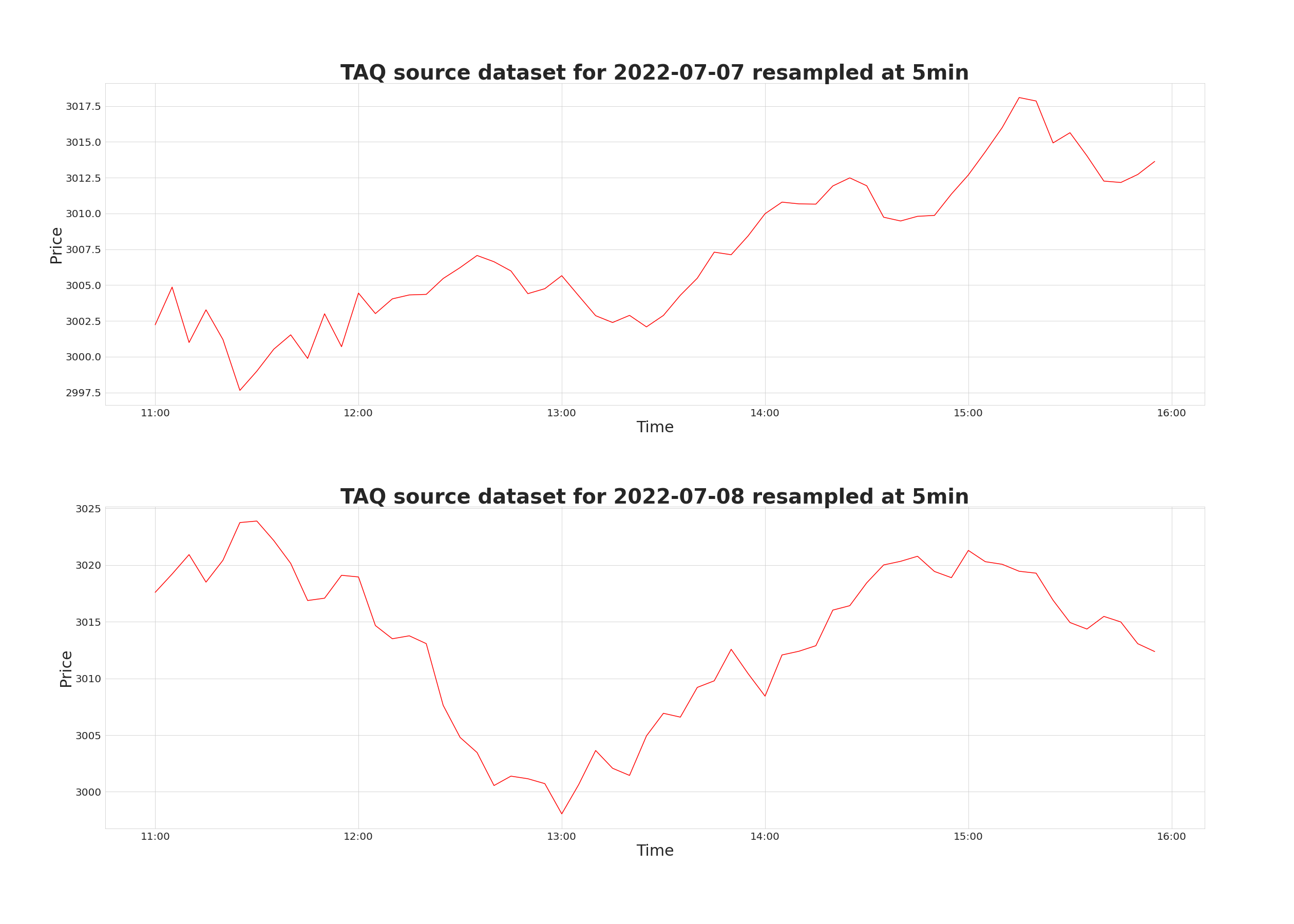}
\caption{Scaled TAQ training data sampled at 5 min interval }
\label{fig:taq_data_scaled}
\end{figure}

Given 1D market time series data how do we extract parameters needed for the construction of GRST(N), i.e. $N$ Gaussian: $G_1 = N(\mu_1,\sigma_1^2)$ at time $t_1$, $G_2 = N(\mu_2,\sigma_2^2)$ at time $t_2$, $\ldots$, $G_N = N(\mu_N,\sigma_N^2)$ at time $t_N$? Specifically, we are interested in constructing GRST(N) tree to model the Brownian Motion represented by a Wiener process projected to a window of time into the future, where the window size is $N$ using past data paralleling GARCH volatility modelling. We used Warton Research Data Services (WRDS) NYSE Trade and Quote (TAQ) dataset \cite{wrds2024} to obtain the training data. The TAQ dataset provides detailed tick-by-tick trade and quote data encompassing all transactions in the U.S. National Market System and delivers transaction data on an intraday basis, down to the microsecond, spanning more than 10,000 stock issues across 16 major American exchanges such as the New York Stock Exchange (NYSE), American Stock Exchange (AMEX), and Nasdaq. The data wrangling task starts with the visualization of the raw data, after initially filtering the dataset on 11:00-16:00 trading times and adjusted sampling to 5 min intervals. 

Scaling is necessary because we are treating each 5 min data aggregate as a sample, however throughout our development the samples are days, so we must "stretch" 5 min interval to one day. Geometric Browning Motion is a good stochastic model for the stock market, which we can clearly see from our sampled raw data as well. Due to self-similarity of the market data, which in our case means that the 5 min interval of a Wiener process path can be extended to an equally valid Wiener process path over one day and can be scaled as \( W_{\text{scale}}(t) = cW(t/c^2) \), for fixed \( c > 0 \). In our case the correct scale factor is $c=\sqrt{N}$, where $N=(1+4)(60/5)=60$ is the number of samples, which is the number of 5 min intervals after sampling raw TAQ data taken during 11:00-16:00 trading times. Figure \ref{fig:taq_data_scaled} shows the results after the scaling have been performed.

We can now extract a feature matrix $Y$ using rolling window statistics. We select future window of size $N$ for which we are trying to model a Wiener path. We now extract the column $Mx1$ feature vector $Y^0_0 = $ 
$\begin{bmatrix}
\alpha_{00} \\
\vdots \\
\alpha_{0M} \\
\end{bmatrix}$
by starting with initial window of size 1 and sliding it across our scaled dataset for day 0 with step one while taking the differences of the corresponding end elements, where $M$ is the resulting size after the sliding window has reached the end of the dataset. We then incrementally adjust the window size and repeat the process  for a feature vector feature vector $Y_j = $ 
$\begin{bmatrix}
\alpha_{j0} \\
\vdots \\
\alpha_{jM} \\
\end{bmatrix}.$
Then feature vector $Y_{M \times N}^0$ is obtained by stacking $Y_j$: \[
Y^0 = \begin{bmatrix}
Y_0 & Y_1 & \cdots & Y_j \cdots & Y_N
\end{bmatrix}.
\]
We repeat the same process for day 1 to construct the feature vector $Y_{M \times N}^1$
\[
Y^1 = \begin{bmatrix}
Y_0 & Y_1 & \cdots & Y_j \cdots & Y_N
\end{bmatrix}.
\]
and continue the process $p$ times where $p$ is the number of training days. Final feature matrix $Y_{pM \times N}$ is constructed by stacking block matrices $Y^i$:
\[
Y = \begin{bmatrix}
Y^0 \\
Y^1 \\
\vdots \\
Y^p
\end{bmatrix}
\]
where each row of $Y$ is a potential Wiener path the market took during $N$ days using $p$ training days of data. The fact that rows of the training data $Y$ are consecutive discrete steps in a Brownian motion is critical when it comes to fitting the Gaussian Mixture model in order to extract the parameters for GRST(N).

\subsection*{GRST(N) Mixture}

\begin{figure} [htbp]
    \centering

\begin{tikzpicture} [ 
declare function={
            normal(\x,\m,\s) = 1/(2*\s*sqrt(pi))*exp(-(\x-\m)^2/(2*\s^2));
        },
    declare function={invgauss(\a,\b) = sqrt(-2*ln(\a))*cos(deg(2*pi*\b));}
       ]
\begin{groupplot}[group style={group size=1 by 1},height=16cm,width=16cm,
    domain=0:16,
    zmin=0, zmax=1,
    xmin=0, xmax=3,
    samples=200,
    samples y=0,
    view={40}{30},
    axis lines=middle,
    enlarge y limits=false,
    xtick={0.5,1.5,2.5},
    xmajorgrids,
    xticklabels={},
    ytick=\empty,
    xticklabels={$t_1$, $t_2$, $t_3$},
    ztick=\empty,
    xlabel=$t$, xlabel style={at={(rel axis cs:1,0,0)}, anchor=west},
    ylabel=$S$, ylabel style={at={(rel axis cs:0,1,0)}, anchor=south west},
    zlabel=Probability density, zlabel style={at={(rel axis cs:0,0,0.5)}, rotate=90, anchor=south},
    set layers
  ]
\pgfplotsinvokeforeach{0.5,1.5,2.5}
{\nextgroupplot[]
  \addplot3 [samples=2, samples y=0, domain=0:2.5] (x, {3*x+2.25}, 0);
  \addplot3 [samples=2, samples y=0, domain=0:2.5] (x, {1.5*(x-0.5)+3}, 0);

  \addplot3 [draw=none, fill=black, opacity=0.25, only marks, mark=dot, mark
  layer=like plot, samples=30, domain=0.1:2.9, on layer=axis background,overlay]
  (#1, {3*(#1)+2.25+invgauss(rnd,rnd)*#1}, 0);  

  \addplot3 [draw=none, fill=black, opacity=0.25, only marks, mark=dot, mark
  layer=like plot, samples=30, domain=0.1:2.9, on layer=axis background,overlay]
  (#1, {1.5*(#1-0.5)+3+invgauss(rnd,rnd)*#1}, 0);

  \begin{pgfonlayer}{axis background}
  \begin{scope}
  \clip (0,0,0) -- (#1,0,0)  -- (#1,14,0) -- (0,14,0) -- cycle;
  \draw[red,no marks,name path=rp1] plot coordinates {\mytabone};
  \draw[red,no marks,name path=rp2] plot coordinates {\mytabtwo};
  \draw[red,no marks,name path=rp3] plot coordinates {\mytabthree};
  \draw[red,no marks,name path=rp4] plot coordinates {\mytabfour};

  \draw[cyan,no marks,name path=rp1b] plot coordinates {\mytaboneB};
  \draw[cyan,no marks,name path=rp2b] plot coordinates {\mytabtwoB};
  \draw[cyan,no marks,name path=rp3b] plot coordinates {\mytabthreeB};
  \draw[cyan,no marks,name path=rp4b] plot coordinates {\mytabfourB};
  
  \end{scope}
  \draw [gray, on layer=axis background] (#1, 2.25+1.5*#1, 0) -- 
  (#1, 2.25+1.5*#1, {normal(0,0,0.5*#1+0.25)});
  \draw [gray, on layer=axis background,name path=vert] (#1,0,0) -- (#1,14,0);

  \draw [gray, on layer=axis background] (#1, 2.25+3*#1, 0) -- 
  (#1, 2.25+3*#1, {normal(0,0,0.85*#1+0.25)});
  \end{pgfonlayer}
  
  \path[name intersections={of=vert and rp1,by={x1}},
  name intersections={of=vert and rp2,by={x2}},
  name intersections={of=vert and rp3,by={x3}},
  name intersections={of=vert and rp4,by={x4}}];
  \draw [fill=red, opacity=0.75, only marks, mark=dot] plot coordinates 
  {(x1) (x2) (x3) (x4)};
  \addplot3 [blue, very thick] (#1, x, {normal(x, 2.25+1.5*#1,0.5*#1+0.25)});

  \path[name intersections={of=vert and rp1b,by={x1b}},
  name intersections={of=vert and rp2b,by={x2b}},
  name intersections={of=vert and rp3b,by={x3b}},
  name intersections={of=vert and rp4b,by={x4b}}];
  \draw [fill=pink, opacity=0.75, only marks, mark=dot] plot coordinates 
  {(x1b) (x2b) (x3b) (x4b)};
  \addplot3 [green, very thick] (#1, x, {normal(x, 2.25+3*#1,0.85*#1+0.25)});
  }
\end{groupplot}

\end{tikzpicture}

\caption{Simulation of the evolution of the Gaussian Mixture Model for the market price with $K=2$ components. The red and cyan Wiener paths represent the Brownian motions for the first component and second components respectively. Black blobs at times $t_1, t_2, t_3$ are samples from corresponding Gaussian mixtures at those times. Red and cyan blobs display the corresponding Wiener paths as it relates to the overall samples from corresponding Gaussian Mixtures (Black blobs). Blue and green Gaussian at times $t_1, t_2, t_3$ represent fitted Gaussian mixture where the mixture probabilities are unchanged throughout the evolution.}
    \label{fig:gauss-mixture-evolut}
\end{figure}
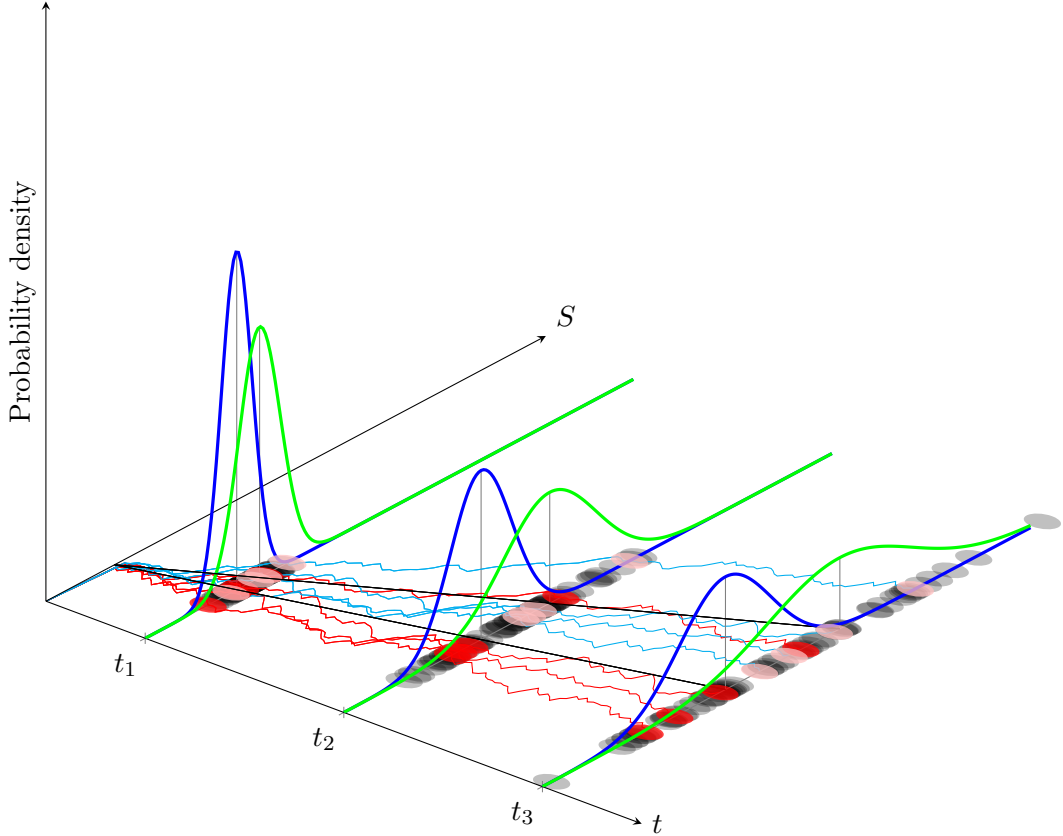

We have transformed 1D time series market data into $N$ dimensional feature matrix $Y$, where $N$ is the size of the future window for which we are trying to model Brownian motion's path a market could take based on historical data. Also, we have a unique training data property stating that each row of $Y$ represents consecutive daily discrete steps in a Brownian motion. We carefully examined what happens to a single Gaussian as it evolves from day 1 to day $N$. Assume we have $K$ Gaussians on day 1 with Gaussian Mixture probabilities $\Pi = \pi_1 \cdots \pi_K$. It is reasonable to model the evolution of each of them independently while keeping original Mixture probabilities $\Pi$ unchanged as shown in Figure \ref{fig:gauss-mixture-evolut}. 

\begin{figure} [htbp]
\centering
\includegraphics[scale=0.2]{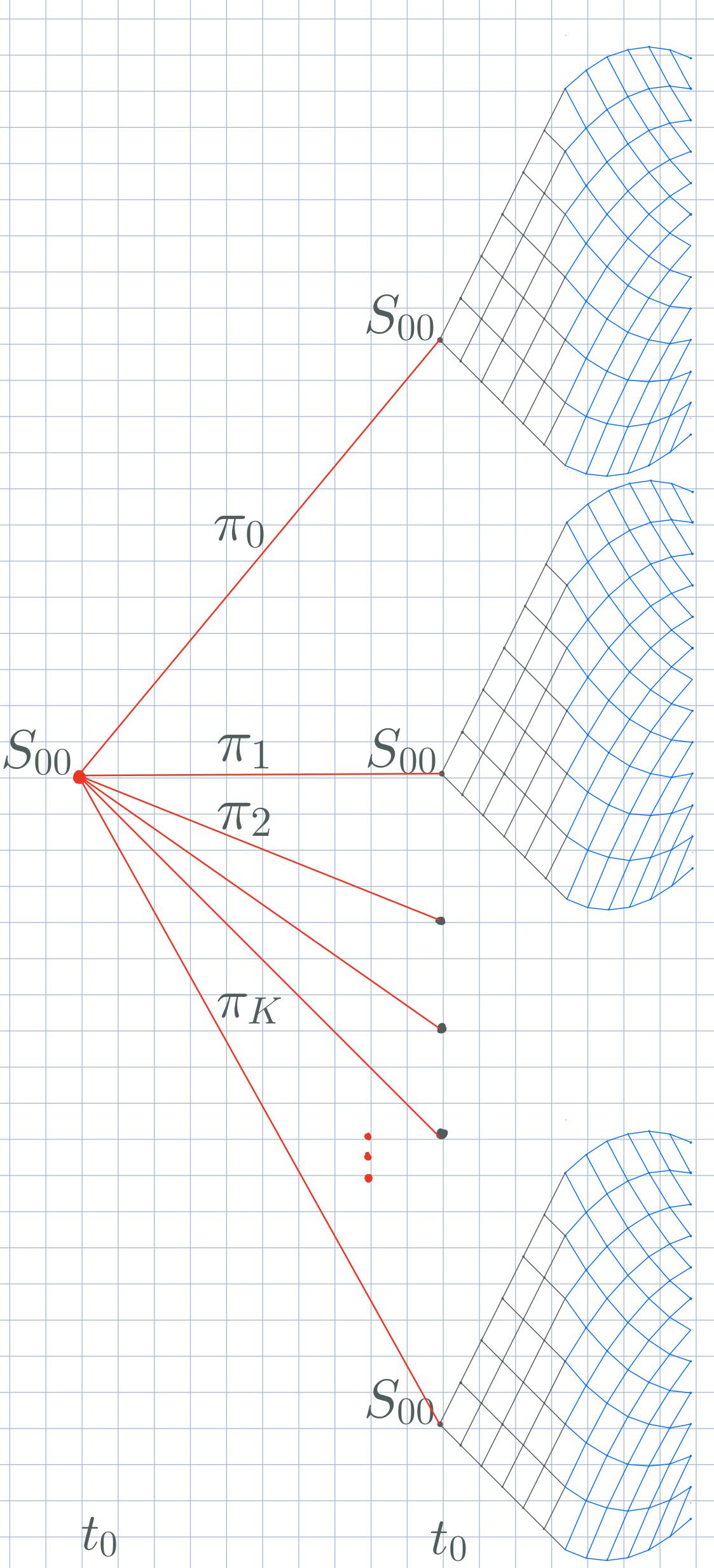}
\caption{GRST(N) Mixture constructed from $K$ GRST(N) trees connected with Gaussian Mixture component probabilities $\Pi$}
\label{fig:grstN-mixture}
\end{figure}

So we use Expectation Maximization for the $N$ dimensional Gaussian Mixture Model to fit $N$ dimensional Gaussian Mixture model and thus obtain the parameters for each of the Gaussians needed for the construction of $K$ GRST(N) trees. Finally, \textbf{GRST(N) Mixture} is defined by a single tree constructed from $K$ GRST(N) trees connected at the root with component mixture probabilities $\Pi$ and initial data value $S_{00}$ transferred unchanged to the roots each of $K$ GRST(N) as shown in Figure \ref{fig:grstN-mixture}. 

\subsection*{Option Pricing}

We can use GRST(N) mixture to price an option using delta-hedging \cite{hull93} on each of $K$ GRST(N) sub-tree until we reach the root of the sub-tree with computed option price of $f(S_i)$ where $i=1 \cdots K$. Using the Fundamental Theorem of Asset Pricing \cite{schachermayer2008freelunch} we employ the expectation view of Risk-Neutral Measure to compute the final price $f$ of the derivative as seen in Equation \ref{eq:grst_pricing}.

\begin{equation} \label{eq:grst_pricing}
    f = \E{f} =  \sum_{i=1}^K  \pi_i f(S_i)
\end{equation}

\section{Conclusion}

This paper introduces a new tree: Gaussian Recombining Split Tree (GRST). It shows careful constructions of GRST and fitting it to the market data. It discusses how to construct GRST Mixture which allows for a much more flexible model of a time series data. Finally, it shows how to price an option using fitted GRST Mixture. We ran back testing for 10 years of market data and found very close agreement in option prices. More work is needed to understand why full covariance matrix is not needed, since we only use variances for construction and how to capture the stochastic process with GRST represents.

\end{document}